\documentclass[aps,twocolumn,groupedaddress]{revtex4}
\usepackage[dvips]{color}
\usepackage{graphicx,color}
\begin{document}

\title{Majorana Kramers Doublets in $d_{x^2-y^2}$-wave Superconductors with Rashba Spin-Orbit Coupling}

\author{Chris L. M. Wong, K. T. Law}

\affiliation{Department of Physics, Hong Kong University of Science and Technology, Clear Water Bay, Hong Kong, China}

\begin{abstract}
In this work, we show that a quasi-one-dimensional $d_{x^2-y^2}$-wave superconductor with Rashba spin-orbit coupling is a DIII class, time-reversal invariant, topological superconductor (TS) which supports a Majorana Kramers Doublet (MKD) at each end of the TS. A MKD is a pair of Majorana end states (MESs) protected by time-reversal symmetry (TRS). An external magnetic field breaks TRS and drives the system from DIII to D class in which case a single MES appears at each end of the TS. We show that a MKD induces resonant Andreev reflection with zero bias conductance peak of $4e^2/h$. Experimental realizations of the proposed model are discussed.
\end{abstract}

\pacs{74.78.-w, 71.10.Pm, 74.25.F-}

\maketitle

{\bf\emph{Introduction}}---A Majorana fermion is a real fermion which has only half the degrees of freedom of a usual Dirac fermion.  It was first pointed out by Read and Green [\onlinecite{RG}] that  Majorana fermions exist at the vortex cores of 2D $p_{x}+ip_{y}$ superconductors and these Majorana fermions are non-Abelian particles [\onlinecite{Ivanov}]. Soon after, Kitaev constructed a spinless fermion model and showed that a single Majorana end state (MES) exists at each end of a $p$-wave superconducting wire [\onlinecite{Kitaev00}]. Recently, several groups proposed that effective p-wave superconductors which support MESs can be realized when s-wave pairings are induced in systems with Rashba spin-orbit coupling[\onlinecite{STF, SLTD,LSD,Alicea,ORV, BDRv,PL,LSD2, PL11,KMB}]. It is predicted that these MESs induce resonant Andreev reflection and cause zero bias conductance (ZBC) peaks in tunneling experiments [\onlinecite{LLN, WADB}]. Remarkably, these ZBC peaks are observed in recent experiments [\onlinecite{Mourik, Deng, Das}], even though the origin of these ZBC peaks is still under hot debate [\onlinecite{Lin, Stanescu,Pientka, LPLL,Pikulin}].

To realize MESs in semi-conductor wires with s-wave superconducting pairing,  an external magnetic field is needed to break the Kramers degeneracy. MESs appear in the regime where an odd number of transverse subbands are occupied [\onlinecite{LSD, Alicea,ORV, BDRv,PL,LSD2,PL11,KMB}]. In this work, we show that Majorana Kramers doublets (MKDs) can be realized in quasi-one-dimensional wires with Rashba coupling and $d_{x^2-y^2}$-wave pairing, in the \emph{absence} of a magnetic field. A MKD is a pair of MESs localized at one end of the wire which is protected by time-reversal symmetry (TRS) [\onlinecite{TK}].

\begin{figure}
\includegraphics[width=3.2in]{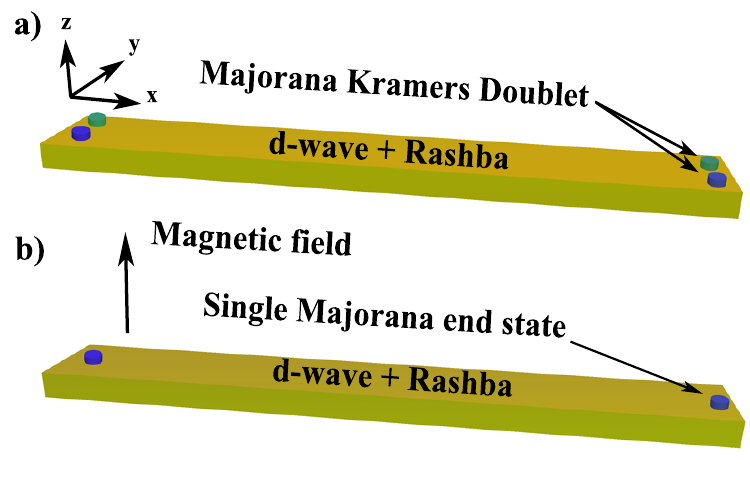}
\caption{\label{picture} A quasi-one dimensional $d_{x^2-y^2}$-wave superconductor with spin-orbit coupling.  a) Majorana Kramers Doublets appear in the absence of an external magnetic field. b) Single MESs appear in the presence of a magnetic field. }
\end{figure}

According to symmetry classification, without a magnetic field, a quasi-one dimensional superconductor with $d_{x^2-y^2}$-wave pairing and Rashba terms is in DIII class as the system respects TRS, particle-hole symmetry (PHS) and breaks spin-rotation symmetry [\onlinecite{SRFL}].  In the following, we show that in the topologically non-trivial regime, a MKD appears at each end of the wire. Interestingly, an external magnetic field breaks TRS and drives the system from DIII class to D class. In this case, a single MES appears at each end of the wire. The schematic pictures of MKDs and single MESs are depicted in Fig.\ref{picture}.

It is shown previously that a single MES in a TS induces quantized ZBC peak of $2e^2/h$ at zero temperature in tunneling experiments. In this work, we show that a MKD in the DIII class TS induces a quantized ZBC peak of $4 e^2/h$. We suggest that a centro-symmetric $d_{x^2-y^2}$-wave superconductor, $\text{CeCoIn}_{5}$, is a candidate of realizing the proposed DIII class TS given that inversion symmetry on the surfaces is broken.

{\bf \emph{Strictly 1D model}}--- Before studying the more realistic quasi-one-dimensional quantum wires with $d_{x^2-y^2}$-wave pairing, we first consider a strictly one-dimensional version of the proposed DIII class TS. We show that the strictly one-dimensional model supports MKDs in the \emph{absence} of an external magnetic field.

A Hamiltonian which describes a strictly one-dimensional TS and supports MKDs can be written as:
\begin{equation}
\begin{array}{l}
H_{1D}= H_{t} + H_{SO}+H_{SC} +H_{Z} \\
H_{t}= \sum_{j,\alpha} -t (\psi^{\dagger}_{j+1,\alpha}\psi_{j \alpha}+h.c.)-\mu \psi^{\dagger}_{j,\alpha}\psi_{j \alpha}\\
H_{SO}=\sum_{j,\alpha,\beta}-\frac{i}{2} \alpha_{R} \psi^{\dagger}_{j+1,\alpha} (\sigma_{y})_{\alpha,\beta} \psi_{j,\beta}+h.c.\\
H_{SC}=\sum_{j} \frac{1}{2}\Delta_{0} (\psi^{\dagger}_{j+1,\uparrow}\psi^{\dagger}_{j,\downarrow}-\psi^{\dagger}_{j+1,\downarrow}\psi^{\dagger}_{j,\uparrow})+h.c. \\
H_{Z}=\sum_{j} V_{z}(\psi^{\dagger}_{j\uparrow}\psi_{j\uparrow}-\psi^{\dagger}_{j\downarrow} \psi_{j\downarrow}),
\end{array} \label{H1d}
\end{equation}
where $H_{t}$, $H_{SO}$, $H_{SC}$ and $H_{Z}$ are the kinetic, spin-orbit coupling, superconducting pairing and the Zeeman coupling terms of the Hamiltonian respectively. Here, $\psi_{j}$ is a fermion operator at site $j$, $\alpha$ and $\beta$ are the spin indices, $t$ is the hopping amplitude, $\alpha_{R}$ is the spin-orbit coupling strength, $\Delta_{0}$ is the superconducting pairing amplitude, and $\sigma_y$ is a Pauli spin matrix. $V_z$ denotes the strength of the Zeeman term. It is important to note that singlet nearest neighbor pairings are introduced in $H_{1D}$ such that the pairing terms are $\cos k$ dependent in momentum space. This is in sharp contrast to the $k$ independent s-wave (on-site superconducting) pairing introduced in previous works [\onlinecite{LSD, Alicea,ORV, BDRv,PL,LSD2,PL11,KMB}]. 

\begin{figure}
\includegraphics[width=3.4in]{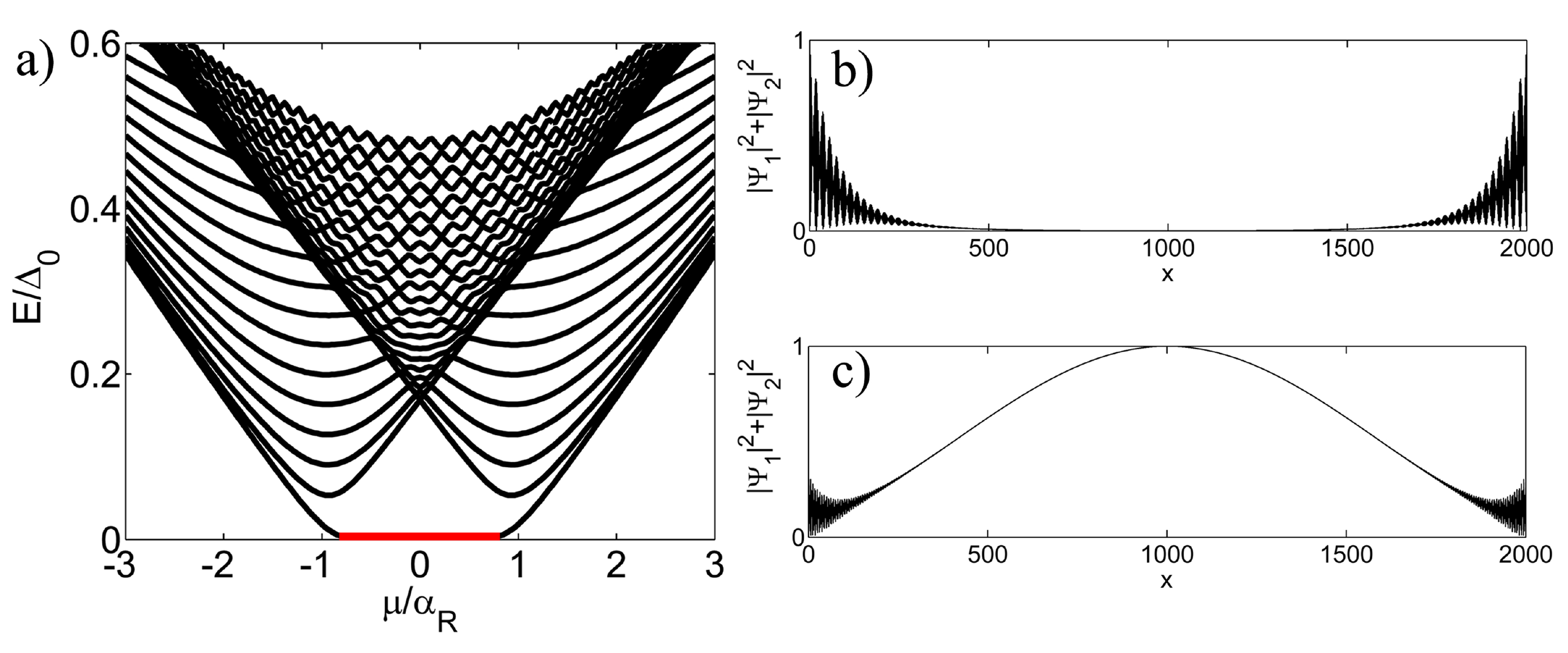}
\caption{\label{BDIspec} a) Excitation energy versus chemical potential. The parameters of $H_{1D}$ are: $L=2000a$, $t=12$, $\Delta_0=1$ and $\alpha_{R}=4$, where $L$ is the length of the wire, $a$ is the lattice spacing. Zero-energy modes exist in the region highlighted in red. b) The sum of the amplitude of the ground state wavefunctions, $|\Psi_{1}|^2+|\Psi_{2}|^2$, versus $x$ where $x$ is the site label. In the topologically non-trivial regime, e.g. $\mu=\alpha_{R}/2$, the ground-state wave functions are localized at the edge. c) In the topologically trivial regime, e.g. $\mu=1.5 \alpha_{R}$, the ground-state wave functions are predominantly in the bulk.}
\end{figure}

The energy spectrum of $H_{1D}$ with $V_z=0$ is shown in Fig.\ref{BDIspec}a. Due to Kramers degeneracy, every state in Fig.2a is doubly degenerate. It is evident from the energy spectrum that zero energy modes exist when the chemical potential satisfies $|\mu| < |\alpha_{R}|$. The sum of the amplitudes of the two ground-state wave functions is shown in Fig.\ref{BDIspec}b to comfirm that the zero energy modes are end states. Since the ground state is doubly degenerate, we expect that there are two MESs, a MKD, at each end of the wire. In the topologically trivial regime where $|\mu| > |\alpha_{R}|$, the ground state wave-functions are predominantly in the bulk as shown in Fig.\ref{BDIspec}c.

To understand the topological origin of the MKDs, we note that Hamiltonian $H_{1D}$ in momentum space can be written as:
\begin{equation}
\begin{array}{l}
H_{1D}(k)=\left(
\begin{array}{cc}
 h(k) & \Delta(k) \\
\Delta^{\dagger}(k) & -h^{T}(-k)
\end{array} \right).
\end{array}   \label{Hk}
\end{equation}
Where $h(k)=(-2t\cos k - \mu) \sigma_{0} + \alpha_R \sin k \sigma_{y}$ and $\Delta(k) = \Delta_{0} \cos k  i\sigma_{y}$. At $V_z =0 $, the Hamiltonian respects TRS such that $T H_{1D}(k) T^{-1}= H_{1D}(-k)$, and PHS such that $P H_{1D}(k) P^{-1}= -H_{1D}(-k)$. Here,  $T=U_{T}K$ and $P= U_{P}K$, where  $ U_{T}=\sigma_0 \otimes i\sigma_y$, $U_{P}=\sigma_x \otimes \sigma_0 $ and $K$ is the complex conjugate operator. Since spin-rotation symmetry is also broken by the Rashba term, $H_{1D}(k)$ is in DIII class. As shown in Appendix A, $H_{1D}(k)$ can be continuously deformed into a flat band Hamiltonian $Q(k)$ which is off diagonalized:
\begin{equation}
Q(k)= \left(
\begin{array}{cc}
 0 & q(k) \\
q^{\dagger}(k) & 0
\end{array} \right),
\end{equation}
where 
\begin{equation}
q(k)=\frac{1}{2} [e^{i\theta_{-}(k)}(\sigma_{0}-\sigma_{y})  + e^{i\theta_{+}(k)}(\sigma_{0}+\sigma_{y}) ],
\end{equation}
and $e^{i\theta_{\pm}(k)}=\frac{-2t\cos(k)-\mu\pm \alpha_{R}\sin(k)+i \Delta_{0}\cos(k)}{\sqrt{[-2t\cos(k)-\mu \pm \alpha_{R}\sin(k)]^2+ [\Delta_{0}\cos(k)]^2}}$.

The DIII class Hamiltonian can be classified by the $Z_2$ topological invariant \cite{TK,QHZ,SR}:
\begin{equation}
N_{DIII}=\frac{\text{Pf}[Tq(k=\pi)]}{\text{Pf}[Tq(k=0)]} \text{exp}\{-\frac{1}{2} \int_{0}^{\pi} dk \text{Tr}[q^{\dagger}(k)\partial_{k}q(k)]\}.
\end{equation}
Here, $\text{Pf}$ denotes the Pfaffian, $T=i\sigma_y$. $N_{DIII}$ can be $1$ or $-1$. The system is topologically trivial when $N_{DIII}=1$. When $N_{DIII}=-1$, the system is in the topologically non-trivial regime and the superconducting wire supports a MKD at each end of the wire as shown in Ref.[\onlinecite{TK}]. For $H_{1D}(k)$, it can be shown that $N_{DIII}=-1$ when $ |\mu|<\alpha_{R}$ and $N_{DIII}=1$ otherwise. This explains the appearance of the zero energy modes in Fig.\ref{BDIspec}a. It is important to note that finite Rashba terms are essential for the appearance of the MKDs. These MKDs are different in origin from the zero energy bounded states of $d_{x^2-y^2}$-wave superconductors with no spin-orbit coupling terms in which case the zero energy modes are not protected against disorder. Moreover, without Rashba terms, the zero energy modes do not appear on surfaces perpendicular to the x-axis [\onlinecite{Hu, Ryu, Kashiwaya}].

It is important to note that the two Majorana fermions of a MKD do not couple to each other due to TRS. Suppose we denote the two Majorana fermions at one end of the wire by $\gamma_1$ and $\gamma_2$ respectively.  We have $T \gamma_1 = \gamma_2$ and $T \gamma_2 = - \gamma_1$, where $T$ is the TRS operator. The extra minus sign in the second equation reflects the fact that $T^2=-1$. If the two Majorana fermions couple to each other, the coupling term can be written as $i \omega \gamma_1 \gamma_2$. However, this coupling term breaks TRS as $ T i \omega \gamma_1 \gamma_2  T^{-1} = - i \omega \gamma_1 \gamma_2$. Therefore, the coupling between the to Majorana fermions of a MKD is not allowed as long as TRS is preserved. This is very different from the case of a D-class superconducting wire in which two Majorana end states at one end of the wire always couple to each other to form a finite energy end state.

To further verify the claim of having two MESs at each end of the wire, we note that $H_{1D}$ can be block diagonalized by a unitary transformation such that
\begin{equation}
U^{-1}H_{1D} U = \left(
\begin{array}{cc}
 H_{+} & 0 \\
0 & H_{-}
\end{array} \right),
\end{equation}
where $H_{\pm}=-(2t\cos k + \mu  \pm \alpha_{R} \sin k)\sigma_{z} + \Delta_{0} \cos k \sigma_{y}$. It is interesting to note that $H_{\pm}$ respect the chiral symmetry $\sigma_{x} H_{\pm} \sigma_{x}=- H_{\pm}$ such that the $H_{\pm}$ are in the AIII class. In the basis which diagonalize $\sigma_{x}$, $H_{\pm}$ can be off-diagonalized as:
\begin{equation}
\tilde{H}_{\pm} = \left(
\begin{array}{cc}
0  & q_{ \pm}(k) \\
q_{\pm}^{\dagger} (k) &  0
\end{array} \right),
\end{equation}
where $q_{\pm}(k)=-(2t \cos k +\mu ) \mp \alpha_{R} \sin k + i \Delta_{0} \cos k$. Define $A_{\pm}(k) = e^{i \phi_{\pm} (k)}= q_{\pm}(k)/|q_{\pm}(k)| $, the AIII class Hamiltonians can be classified by the topological invariant
\begin{equation}
N_{AIII}^{\pm}= \frac{1}{2\pi} \int_{-\pi}^{\pi}\frac{dA_{\pm}(k)}{A_{\pm}(k)}.
\end{equation}
Non-zero $N_{AIII}$ indicates the appearance of MESs. It can be shown that $|N_{AIII}^{\pm}|=1$ when $|\mu| < |\alpha_{R}|$ and $|N_{AIII}^{\pm}|=0$ otherwise. Since both $H_{+}$ and $H_{-}$ are topologically non-trivial in the regime $|\mu| < |\alpha_{R}|$, there are two  MESs in the corresponding regime. This is analogous to a time-reversal invariant 2D $p \pm ip$ superconductor which can be regarded as two copies of spinless chiral p-wave superconductors and each copy of the chiral p-wave superconductor can be classified by Chern numbers.

\begin{figure}
\includegraphics[width=3.4in]{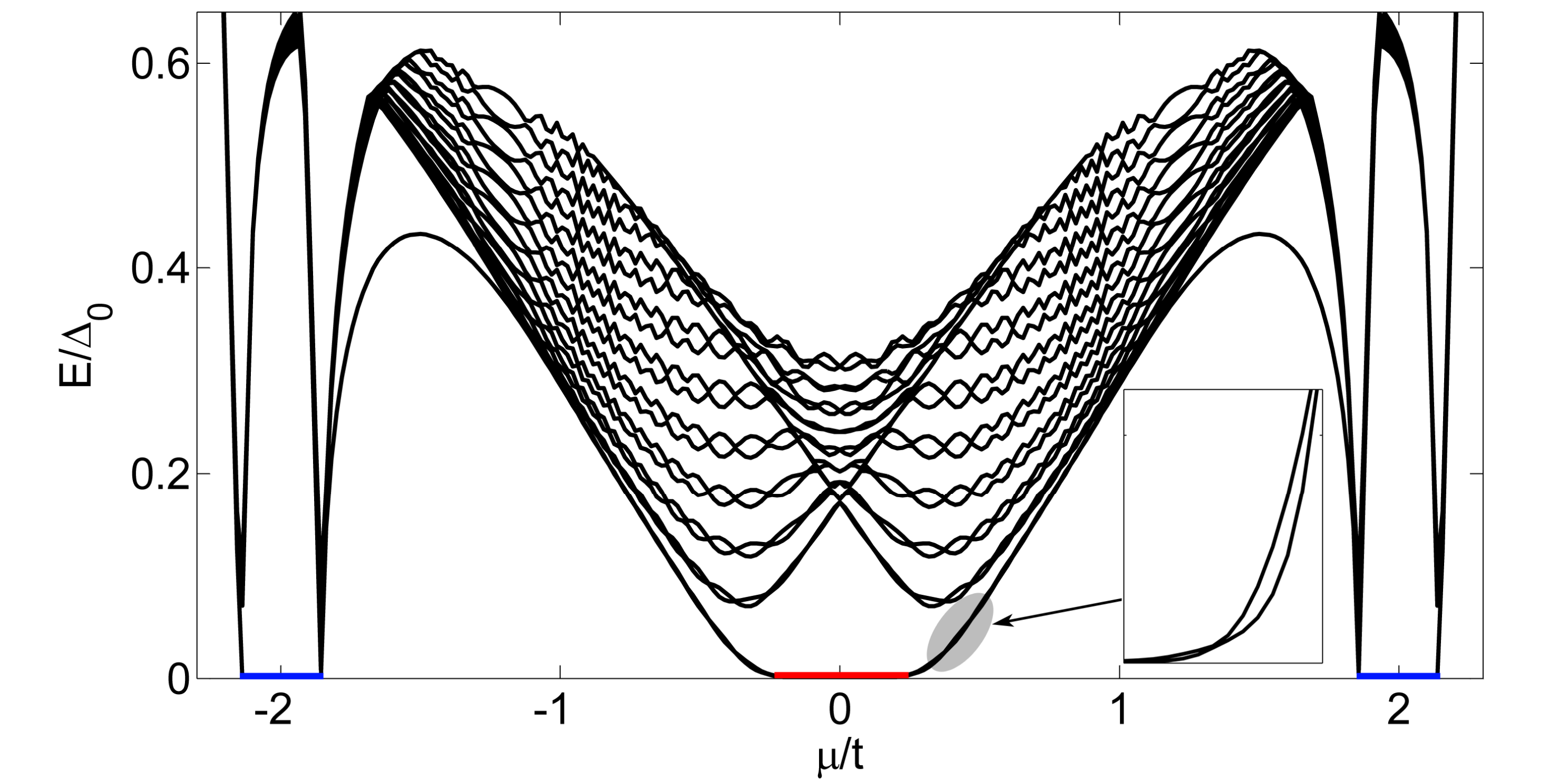}
\caption{\label{BDIVz} Excitation energy as a function of chemical potential. The parameters of $H_{1D}$ are: $t=12$, $\Delta_0=1$, $\alpha_R=4$ and $V_z=2$. Single MESs appear near the band bottom at $\mu \approx -2t$, indicated by the blue line. Double MESs appear near the band center at $\mu \approx 0$, indicated by the red line. The insert is an enlargement of the gray section which indicates that there are two states near zero energy at $\mu \approx 0$. }
\end{figure}

{\bf \emph{Strictly 1D model with finite $V_z$}}--- When $V_z$ is non-zero, TRS is broken and the Hamiltonian is no longer in DIII class. In this section, we show that MESs appear even in the presence of an external magnetic field. The energy eigenvalues of $H_{1D}$ with finite $V_z$ versus the chemical potential are shown in Fig.\ref{BDIVz}. It is interesting to note that the zero energy modes appear in two separate regimes. In the regime near the band bottom, $\mu \approx -2t$, there is a single zero energy mode in the excitation spectrum which corresponds to a single MES at each end of the wire.

Remarkably, two nearly zero energy modes appear near the middle of the band, at $\mu \approx 0$, which correspond to two MESs at each end of the wire. It is shown in Appendix B that the double MESs are stable even in the presence of disorder. This is in sharp contrast to a one-dimensional TS in D class in which only a single MES at each end of the wire is stable [\onlinecite{LSD, Alicea,ORV, BDRv,PL,LSD2,PL11,KMB}]. 

To understand the appearance of single and double MESs at different chemical potential, we note that the strictly one-dimensional system with non-zero $V_z$ is in the BDI class. This is because the Hamiltonian respects a time-reversal like symmetry $T_{BDI}$ such that $T_{BDI} H_{1D}(k) T_{BDI}^{-1}= H_{1D}(-k)$ even though TRS is broken. Here, $T_{BDI}=\sigma_{0}\otimes \sigma_{0} K$ and $T_{BDI}^{2}=1$. Since PHS is respected as before, the Hamiltonian with finite $V_{z}$ is in the BDI class. Since one-dimensional systems in BDI class are classified by integers, therefore, it is possible to have multiple stable MESs at the end of the wire [\onlinecite{SRFL,TK}].

In Appdendix B, we show that $H_{1D}$ with finite $V_{z}$ can be classified by a topological invariant $N_{BDI}$ that $|N_{BDI}|=1$ in the regime where single MESs appear, e.g., when $\mu \approx -2t$. On the other hand, $N_{BDI}=2$ near $\mu \approx 0$ where two MESs appear. The condition for $N_{BDI}=-1$ is:
\begin{equation}
(2t + \mu)^2 < V_{z}^2 - \Delta_{0}^2.
\end{equation}
This is exactly the same condition for single MESs to appear in the s-wave pairing case [\onlinecite{STF, SLTD}].
The conditions for $N_{BDI}=2$ is:
\begin{equation}
\quad \mu^2  < V_{z}^2+\alpha_{R}^2. \label{BDG2}    \label{BDIN2}
\end{equation} 
It is interesting to note that the conditions for MKDs to appear, $ |\mu| < |\alpha_R|$, is reproduced in Eq.\ref{BDIN2} by setting $V_z=0$. The double MESs at finite $V_z$ can be understood as the descendants of the MKDs.

It is important to note that the BDI classification applies only in the strictly 1D limit, when the symmetry $T_{BDI} H_{1D}(k) T_{BDI}^{-1}= H_{1D}(-k)$ is respected. In quasi-one dimensional case, this symmetry is broken and the Hamiltonian is in the D class in the presence of $V_z$. Therefore, double MESs in the quasi-one dimensional case are not stable if TRS is broken.

{\bf \emph{Multi-channel case}}---In this section, we consider the quasi-one dimensional limit in which multiple transverse sub-bands of a wire are occupied. In the quasi-one-dimensional case, the Hamiltonian can be written as:
\begin{equation}
\begin{array}{ll}
H_{q1D} = & H_{t} + H_{SO}+H_{SC} +H_{Z}, \\
H_{t}= & \sum_{\mathbf{R},\mathbf{d}, \alpha} -t(\psi^{\dagger}_{\mathbf{R+d},\alpha}\psi_{\mathbf{R}, \alpha}+h.c.)-\mu \psi^{\dagger}_{\mathbf{R},\alpha}\psi_{\mathbf{R}, \alpha}\\
H_{SO}= & \sum_{\mathbf{R,d},\alpha,\beta}-\frac{i}{2} \alpha_{R} \psi^{\dagger}_{\mathbf{R+d},\alpha} \hat{\mathbf{z}}\cdot(\vec{\sigma}_{\alpha \beta}\times \mathbf{d})\psi_{\mathbf{R},\beta}+h.c.\\
H_{SC}= & \sum_{\mathbf{R}}\frac{1}{2}[ \Delta_{0} (\psi^{\dagger}_{\mathbf{R+d_{x}},\uparrow}\psi^{\dagger}_{\mathbf{R},\downarrow}-\psi^{\dagger}_{\mathbf{R+d_{x}},\downarrow}\psi^{\dagger}_{\mathbf{R},\uparrow})- \\ 
&\Delta_{0}(\psi^{\dagger}_{\mathbf{R+d_{y}},\uparrow}\psi^{\dagger}_{\mathbf{R},\downarrow}-\psi^{\dagger}_{\mathbf{R+d_{y}},\downarrow}\psi^{\dagger}_{\mathbf{R},\uparrow}) +h.c. ]\\
H_{Z}= & \sum_{\mathbf{R}} V_{z}(\psi^{\dagger}_{\mathbf{R}\uparrow}\psi_{\mathbf{R}\uparrow}-\psi^{\dagger}_{\mathbf{R}\downarrow} \psi_{\mathbf{R}\downarrow}).
\end{array}  \label{Hq1D}
\end{equation}

\begin{figure}
\includegraphics[width=3.2in]{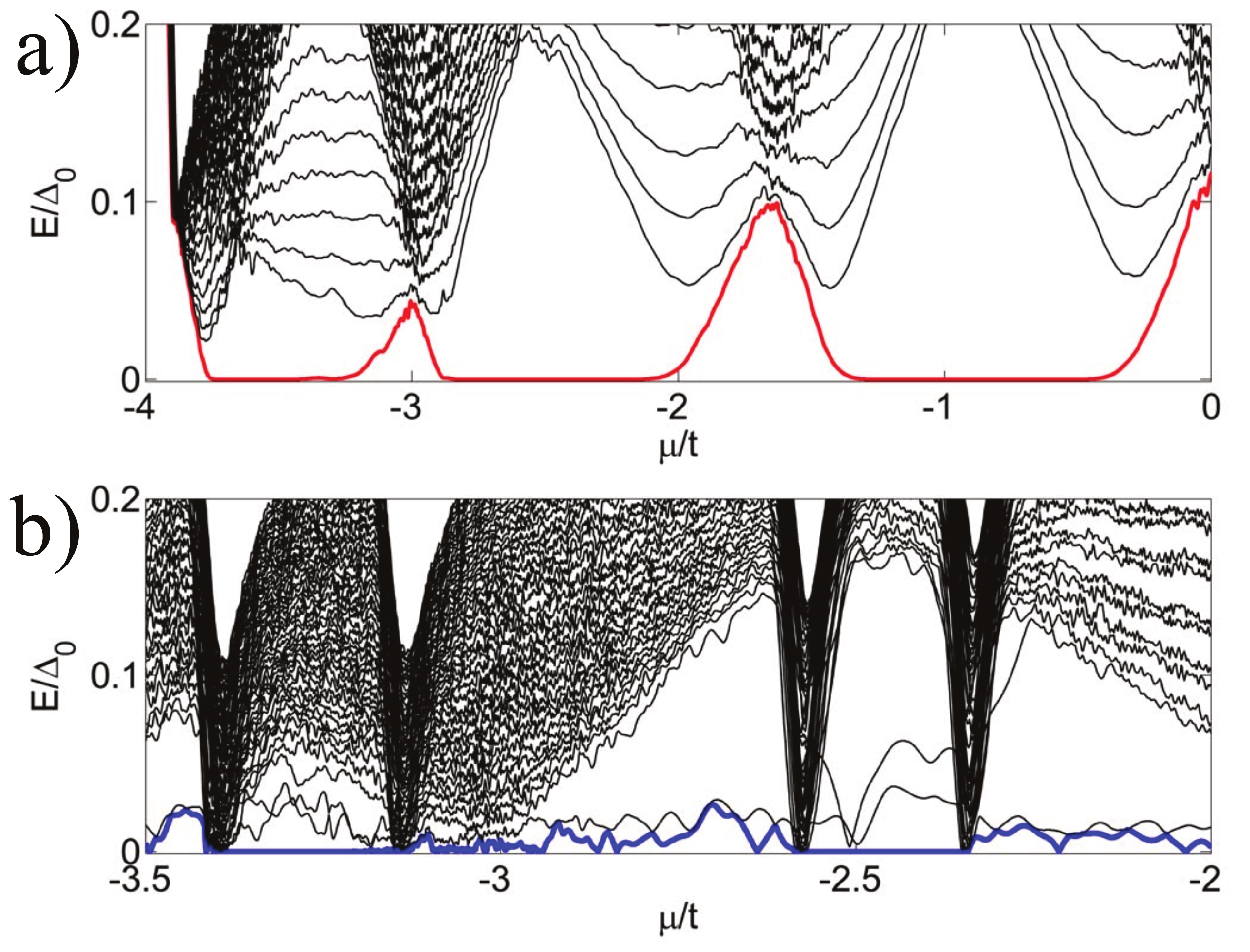}
\caption{\label{Q1D}  a) Excitation energy versus chemical potential. The parameters in $H_{q1D}$ are $V_z=0$, $t=12$, $\Delta_0=1$ and $\alpha_R=4$. On-site Gaussian disorder with vairance $\Delta_{0}^2$ is present. The width of the wire is $W=6a$ and the length is $L=1400a$. All the states shown are doubly degenerate. Zero energy modes associated with MKDs appear in a wide range of chemical potential. The Kramers degenerate ground states are depicted in red. b) The parameters in b) are the same as in a) except $V_z=2$. Non-degenerate zero energy modes appear even when TRS is broken. The ground state is depicted in blue. }
\end{figure}

Here, $\mathbf{R}$ denotes the lattice sites, $\mathbf{d}$ denotes the two unit vectors $\mathbf{d_{x}}$ and $\mathbf{d_y}$ which connects the nearest neighbor sites in the $x$ and $y$ directions respectively. This model is the same as the tight-binding model in Ref.[\onlinecite{PL11}] except for the superconducting pairing terms. The pairing terms in $H_{q1D}$ can be written as $\Delta_{0}[ \cos(k_{x})- \cos(k_{y})]$ in the momentum space. Therefore, $H_{q1D}$ describes a quantum wire with spin-orbit coupling and a $d_{x^2-y^2}$-wave superconducting pairing.

The energy spectrum of $H_{q1D}$ with $V_z =0$ and on-site disorder is shown in Fig.\ref{Q1D}a. The length of the wire is chosen to be much larger than the superconducting coherence length $L \gg t / \Delta_0 $ and the width is on the order of the coherence length. It is evident from Fig.\ref{Q1D}a that the zero energy modes appear for a wide range of chemical potential. Due to Kramers theorem, every state in Fig.\ref{Q1D}a is doubly degenerate and the zero energy modes are associated with MKDs at each end of the sample [\onlinecite{TK}]. It is important to note that the zero modes are robust against disorder which does not break TRS. In principle, the topological invariant of the quasi-one dimensional wire can be calculated using Eq.5. In the quasi-one dimensional regime, $q(k)$ in Eq.5 will be a $2N \times 2N$ matrix where $2N$ is the number of transverse subbands of the wire in the normal state. 

In the presence of the $V_z$ term, time-reversal symmetry is broken and the $H_{q1D}$ is in D class. The resulting energy spectrum of $H_{q1D}$ with $V_z=2$ is shown in Fig.\ref{Q1D}b. The non-degenerate zero energy modes appear for a wide range of chemical potential which are associated with single MESs at the sample end.

{\bf \emph{Resonant Andreev reflection}}--- It is shown previously that a single Majorana fermion induces a quantized ZBC peak of $G=2\frac{e^2}{h}$ in Andreev reflection experiments [\onlinecite{LLN, WADB}] when a normal metal lead couples to a MES. Here, we show that the a MKD in DIII class TS induces a ZBC peak of $G=4\frac{e^2}{h}$ instead. 

\begin{figure}
\includegraphics[width=3.2in]{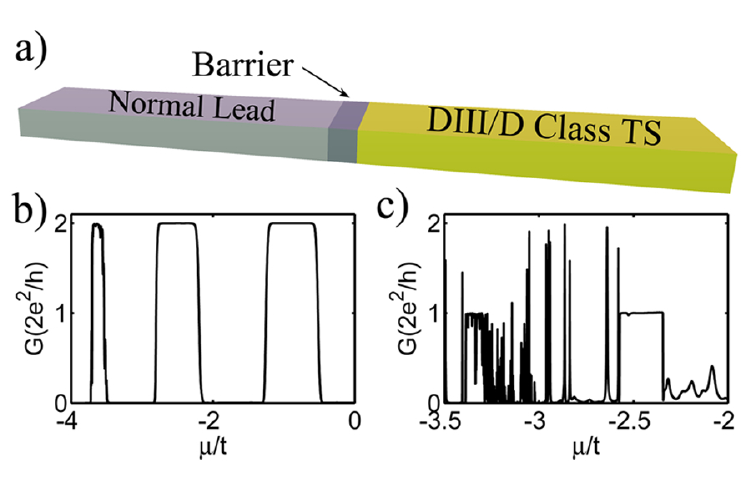}
\caption{\label{AR}  a) A normal lead is attached to the end of a TS. b) ZBC from the normal lead to the TS versus chemical potential at $V_z=0$. The ZBC is quantized at $4e^2/h$ in the topologically non-trivial regime, due to the presence of double MESs. c) ZBC versus chemical potential at $V_z=2\Delta_0$. The ZBC is quantized at $2e^2/h$ in the topologically non-trivial regime due to the presence of a single MES.}
\end{figure}

A TS with parameters given in Fig.\ref{Q1D} is attached to a semi-infinite normal lead as shown in Fig.5a. The hopping amplitudes on the normal lead are the same as the hopping amplitudes on the TS. The barrier is simulated by a reduced hopping matrix element between the TS and the normal lead.  Using lattice Green's function method [\onlinecite{Lee1,Lee2}], the ZBC from a normal lead to a DIII class TS is calculated and shown in Fig.5b.  It is evident that the ZBC is quantized at $4\frac{e^2}{h}$ in the regime where double MESs appear. In the presence of an external magnetic field, the TS is in D class and ZBC is quantized at $2\frac{e^2}{h}$ in the topologically non-trivial regime as shown in Fig.5c.

{\bf \emph{Discussion}}--- A few important comments follow. First, for simplicity, we assumed that the wire is aligned along the x-direction. If a quasi-one dimensional wire is grown along a direction tilted with angle $\theta$ with respect to the x-axis, the superconducting pairing symmetry becomes $ \cos2 \theta ( \cos k_{x'} - \cos k_{y'}) + \sin 2 \theta \sin k_{x'} \sin k_{y'} $. Here, $k_{x'}$ and $k_{y'}$ denote the momenta parallel and perpendicular to the wires respectively. In other words, one obtains a $d_{x^2-y^2} + d_{xy}$-wave pairing superconductor. Since the $d_{xy}$ term does not break the TRS and PHS, the conclusions of this work stand so long as $\theta$ is away from $\pi/4$ or $3\pi/4$ in which directions the pairing gap along the wire vanishes.

Second, only spin singlet $d_{x^2-y^2}$ pairing is considered in the main text. However, in the presence of Rashba terms, spin triplet pairing terms may appear [\onlinecite{GR,Tanaka}]. Nevertheless, spin-triplet terms do not break the TRS and PHS. The presence of spin-triplet terms does not affect the conclusion of this work as long as the bulk gap is not closed by these pairing terms as shown in Appendix C.

Third, the results discussed in this work applies to all quasi-one dimensional $d_{x^2-y^2}$-wave superconductors with Rashba spin-orbit coupling. A candidate material of DIII class TS is a layered heavy fermion superconductor $\text{CeCoIn}_{5}$. Bulk $\text{CeCoIn}_{5}$ is a $d_{x^2-y^2}$-wave superconductor[\onlinecite{Matsuda}]. Unfortunately, due to  inversion symmetry in the bulk, there is no Rashba spin-orbit coupling in the system which is crucial for the topological phases discussed in this paper. However, inversion symmetry is broken at the surface layer such that Rashba spin-orbit coupling terms can be induced on the surface layer as shown by Maruyama et al. [\onlinecite{Sigrist}]. Therefore, the surface layer of a $\text{CeCoIn}_{5}$ thin film can be described by $H_{q1D}$ in Eq.\ref{Hq1D}, with the possibility of having additional triplet pairing terms as discussed above. It is shown in Appendix D that multiple layers of $\text{CeCoIn}_{5}$, with different Rashba strength in different layers, coupled by interlayer hopping terms can support MKDs.

Finally, in Appendix F, we show that MKDs can emerge as end states of a metal wire with Rashba spin-orbit coupling if the wire is placed on top of usual d-wave superconductor without Rashba spin-orbit coupling. Since the d-wave superconductor is nodal, the Majorana end state wavefunctions can leak into the d-wave superconductor. Fortunately, Majorana wavefunctions leak into the nodal directions only and the two MKDs do not couple to each other directly. As a result, the MKDs survive even if the parent superconductor is nodal.

{\bf\emph{Conclusion}}--- We show that quasi-one dimensional $d_{x^2-y^2}$-wave superconductors with Rashba spin-orbit coupling are DIII class TS which support MKDs. Single MESs appear in the presence of a magnetic field. The MKDs induce resonant Andreev reflection with a quantized ZBC peak of $4\frac{e^2}{h}$. We suggest that $\text{CeCoIn}_5$ and metal wires on a d-wave superconductor are candidate materials for this DIII topological superconducting phase.

{\bf\emph{Acknowledgments}}--- We thank C.H. Chung,  C.Y. Hou, P.A. Lee, J. Liu, T.K. Ng, Y. Matsuda, B. Normand and A. Potter for insightful discussions. The authors thank the support of HKRGC through DAG12SC01, Grant 605512 and HKUST3/CRF09.

\appendix
\section{DIII Class Topological Invariant}
In this section, we first obtain the flat band Hamiltonian of $H_{1D}(k)$ in Eq.2 of the main text. We then calculate the topological invariant $N_{DIII}$ defined in Eq.5. 

To obtain the flat band Hamiltonian, we first note that at $V_z=0$, the Hamiltonian $H_{1D}(k)$ in Eq.2 of the main text satisfies time-reversal symmetry  and particle-hole symmetry such that
\begin{equation}
\begin{array}{ll}
U_{T}H_{1D}^{*}(-k)U_{T}^{\dagger}=H_{1D}(k), & \text{and} \\
U_{C}H_{1D}^{*}(-k)U_{C}^{\dagger}=-H_{1D}(k), &
\end{array}
\end{equation}
where  $ U_{T}=\sigma_0 \otimes i\sigma_y$ and $U_{C}=\sigma_x \otimes \sigma_0 $.
As a result of time-reversal symmetry and particle-hole symmetry, the Hamiltonian acquires a chiral symmetry
\begin{equation}
U_{S}^{\dagger}H_{1D}(k)U_{S}=-H_{1D}(k),
\end{equation}
where $U_{S}=i U_{T} U_{C}$. Therefore, $H_{1D}(k)$ is in the DIII class [\onlinecite{SRFL}]. In the basis that $U_{S}$ is diagonal, $H_{1D}(k)$ can be written in the off-diagonal form
\begin{equation}
\tilde{H}_{1D}(k)=VH_{1D}(k)V^{\dagger}=\left(
\begin{array}{cc}
 0 & D(k) \\
D^{\dagger}(k) & 0
\end{array} \right),
\end{equation}
where
\begin{equation}
V=\frac{1}{\sqrt{2}}\left(
\begin{array}{cc}
 \sigma_0 & -\sigma_y \\
\sigma_0 & \sigma_y
\end{array} \right),
\end{equation}
and $D(k)= h(k) + \Delta(k) \sigma_{y} $. Due to the chiral symmetry, the eigenvalues of the Hamiltonian can be written as $\pm \lambda_{a}(k)$ with $a=1,2$. For a gapped Hamiltonian, we can assume $ \lambda_a (k) > 0$ for all $k$.

Let $ (\chi^{\pm}_{a}(k), \eta^{\pm}_{a}(k))^T$ be the eigenfunctions of $\tilde{H}_{1D}(k)$ with eigenvalues $\pm \lambda_{a}(k)$ respectively. Using the eigenvalue equation of $\tilde{H}_{1D}^{2}(k)$, one obtains
\begin{equation}
DD^{\dagger} \chi^{\pm}_{a}(k)=\lambda_{a}^2 \chi^{\pm}_{a}(k), \quad D^{\dagger}D \eta^{\pm}_{a}(k)=\lambda_{a}^2 \eta^{\pm}_{a}(k).
\end{equation}
Therefore, the eigenfunctions of $\tilde{H}_{1D}(k)$ are
\begin{equation}
|\Psi_{a},\pm \rangle =
\left(
\begin{array}{c}
\chi^{\pm}_{a}  \\
\eta^{\pm}_{a}
\end{array} \right) = \frac{1}{\sqrt{2}}
\left(
\begin{array}{c}
u_{a}  \\
\pm D^{\dagger} u_{a} / \lambda_{a}
\end{array} \right),
\end{equation}
where $ u_{a}$ are the normalized eigenfunctions of $D D^{\dagger}$.
Once the wavefunctions are known, we can calculate the flat band Hamiltonian of $H_{1D}(k)$, which is defined as $ Q(k)= \sum_{a=1,2} | \Psi_{a}, + \rangle \langle \Psi_{a},+| - | \Psi_{a},- \rangle \langle \Psi_{a}, -|$. In terms of $u_a$, we have
\begin{equation}
Q(k)= \left(
\begin{array}{cc}
 0 & q(k) \\
q^{\dagger}(k) & 0
\end{array} \right)
=
\sum_{a=1,2} \left(
\begin{array}{cc}
 0 & u_{a}u_{a}^{\dagger} \frac{D(k)}{\lambda_{a}} \\
\frac{D^{\dagger}(k)}{\lambda_{a}} u_{a}u_{a}^{\dagger} & 0
\end{array} \right).
\end{equation}
Using $DD^{\dagger}=[(2 t\cos k+\mu)^2+\Delta_0^2 \cos^2 k + \alpha_R^2 \sin^2 k]\sigma_0- 2\alpha_{R} \sin k(2 t\cos k+\mu) \sigma_{y} $, we have 
\begin{equation}
q(k)=\frac{1}{2} [e^{i\theta_{-}(k)}(\sigma_{0}-\sigma_{y})  + e^{i\theta_{+}(k)}(\sigma_{0}+\sigma_{y}) ],
\end{equation}
where $e^{i\theta_{\pm}(k)}=\frac{-2t\cos(k)-\mu\pm \alpha_{R}\sin(k)+i \Delta_{0}\cos(k)}{\sqrt{[-2t\cos(k)-\mu \pm \alpha_{R}\sin(k)]^2+ [\Delta_{0}\cos(k)]^2}}$. According to Refs.[\onlinecite{QHZ,SR,TK}], the $Z_2$ topological invariant of the system can be written as:
\begin{equation}
N_{DIII}=\frac{\text{Pf}[Tq(k=\pi)]}{\text{Pf}[Tq(k=0)]} \text{exp}\{-\frac{1}{2} \int_{0}^{\pi} dk \text{Tr}[q^{\dagger}(k)\partial_{k}q(k)]\}, 
\end{equation}
where $T=i\sigma_y$. $N_{DIII}$ can be $1$ or $-1$. The system is topologically trivial when $N_{DIII}=1$. When $N_{DIII}=-1$, the system is in the topologically non-trivial regime and the superconducting wire supports a Majorana Kramers Doublet at each end of the wire. For $H_{1D}(k)$ in Eq.2 of the main text, it can be shown that $N_{DIII}=-1$ when $ |\mu|<\alpha_{R}$ and $N_{DIII}=1$ otherwise.


\section{BDI Class Topological Invariant}
In the presence of the $V_z$ terms in $H_{1D}(k)$ of the main text, time-reversal symmetry is broken and one cannot use the $Z_2$ invariant $N_{DIII}$ mentioned in the above section to characterize the Hamiltonian. However, we note that in the strictly one-dimensional case, the Hamiltonian $H_{1D}(k)$ respects a time-reversal like symmetry $T_{BDI} H_{1D}(k) T_{BDI}^{-1}= H_{1D}(-k)$, where $T_{BDI}=\sigma_{0}\otimes \sigma_{0} K$. It is important to note that $T_{BDI}^2=1$. Together with the fact that $H_{1D}(k)$ respects the particle-hole symmetry $P H_{1D}(k) P^{-1}= -H_{1D}(-k)$ as before, with $P^2=1$, $H_{1D}(k)$ is in the BDI class [\onlinecite{SRFL}].

It is well known that BDI class Hamiltonians in one dimension are classified by integer numbers \cite{SRFL}. In this section, we show how the integer topological invariant can be calculated following Ref.\onlinecite{TS}.

\begin{figure}
\includegraphics[width=3.4in]{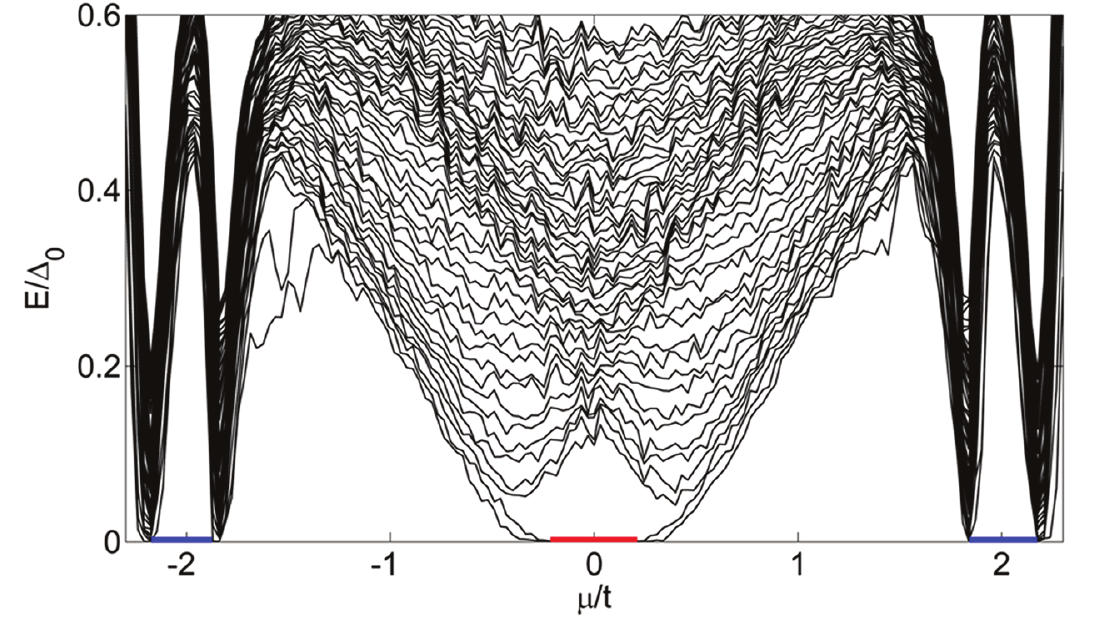}
\caption{\label{SF1} Excitation energy of $H_{1D}(k)$ as a function of chemical potential in the presence of disorder. The parameters in this figure are the same as the  parameters in Fig.3 of the main text except the fact that on-site potential disorder with variant $\Delta_{0}^2$ is added to the Hamiltonian. This is in sharp contrast to the multi-channel case when the system is in the D class in which case an even number of Majorana fermions are not protected against disorder.}
\end{figure}

Due to the $T_{BDI}$ symmetry and the particle-hole symmetry, the Hamiltonian $H_{1D}(k)$ acquires a chiral symmetry $S=\sigma_{x} \otimes \sigma_0 $ such that $S H_{1D}(k) S^{-1} = - H_{1D}(k)$. In the basis that $S$ is diagonal, the Hamiltonian can be written in the off-diagonal form
\begin{equation}
W H_{1D}(k) W^{\dagger}= \left(
\begin{array}{cc}
 0 & A(k) \\
A^{T}(-k) & 0
\end{array} \right),
\end{equation}
where
\begin{equation}
W=\frac{1}{\sqrt{2}}\left(
\begin{array}{cc}
 \sigma_{x} & -\sigma_{x} \\
\sigma_{x} & \sigma_{x}
\end{array} \right), \quad \text{and} 
\end{equation}
\begin{equation}
 A(k)= \left(
\begin{array}{cc}
 -2t\cos k-\mu -V_z & i \alpha_R \sin k - \Delta_0 \cos k \\
 -i \alpha_R \sin k + \Delta_0 \cos k &  -2t \cos k -\mu +V_z
\end{array} \right).
\end{equation}
Note that $A(k)$ is real at $k=0, \pm \pi$, we can define the quantity
\begin{equation}
z(k)=e^{i \theta(k)}=\text{Det}[A(k)]/|\text{Det}[A(k)]|,
\end{equation}
such that $\theta(k) = n \pi$ at $k=0, \pm \pi$ with integer $n$. The winding number of $\theta(k)$ can be used as the topological invariant which characterizes the Hamiltonian $H_{1D}(k)$. The winding number $N_{BDI}$ can be written as
\begin{equation}
N_{BDI}=\frac{-i}{\pi} \int_{k=0}^{k=\pi} \frac{dz(k)}{z(k)}.
\end{equation}
It counts the number of Majorana end states at one end of a superconducting wire \cite{TS,EG}. Using $A(k)$ obtained from $H_{1D}(k)$, it can be easily shown that $N_{BDI}=1$ when
\begin{equation}
(2t - \mu)^2 < V_{z}^2 - \Delta_{0}^2 \quad \text{and} \quad (2t + \mu)^2 > V_{z}^2 - \Delta_{0}^2.
\end{equation}
$N_{BDI}=-1$ when
\begin{equation}
(2t + \mu)^2 < V_{z}^2 - \Delta_{0}^2 \quad \text{and} \quad (2t - \mu)^2 > V_{z}^2 - \Delta_{0}^2.
\end{equation}
$N_{BDI}=2$ when
\begin{equation}
(2t \pm \mu)^2 > V_{z}^2 - \Delta_{0}^2 \quad \text{and} \quad \mu^2  < V_{z}^2+\alpha_{R}^2. \label{BDG2}
\end{equation}
Assuming $2t \gg |V_z|$ and $\Delta_0$, we have the Eq.{9} and Eq.{10} of the main text. It is interesting to note that when $V_Z$ satisfies Eq.\ref{BDG2}, there are two Majorana end states at each end of the superconducting wire even when time-reversal symmetry is broken. These double Majorana end states are topologically protected and they survive in the presence of disorder as shown in Fig.\ref{SF1}.

\section{The effect of spin singlet and spin-triplet pairing terms}
In the main text, a quasi-one dimensional superconductor with a pure $d_{x^2-y^2}$-wave pairing and Rashba spin-orbit coupling is studied. The Hamiltonian $H_{q1D}$ is in the DIII class which may support Majorana Kramers Doublets in the absence of an external magnetic field. However,  due to the presence of the Rashba terms, additional s-wave spin-singlet and p-wave single-triplet pairing channels may exist. In this section, we show that the proposed topological state is stable in the presence of the s-wave and p-wave pairing channels.

In this section, s-wave spin-singlet and p-wave spin-triplet pairing terms are added to $H_{q1D}$ in Eq.11 of the main text. These pairing terms can be written as:
\begin{equation}
\begin{array}{ll}
H_{\Delta_s} = & \Delta_s \sum_{\mathbf{R}} \psi^{\dagger}_{\mathbf{R},\uparrow} \psi^{\dagger}_{\mathbf{R}, \downarrow} + h.c. \\
H_{\Delta_p} = & \frac{1}{2} \Delta_{p} \sum_{\mathbf{R},\sigma}  [(\psi^{\dagger}_{\mathbf{R}+\mathbf{d_x},\sigma}\psi^{\dagger}_{\mathbf{R},\sigma}-\psi^{\dagger}_{\mathbf{R}-\mathbf{d_x},\sigma} \psi^{\dagger}_{\mathbf{R},\sigma})  \\ 
&   -i\epsilon_{\sigma} (\psi^{\dagger}_{\mathbf{R}+\mathbf{d_y},\sigma}\psi^{\dagger}_{\mathbf{R},\sigma}-\psi^{\dagger}_{\mathbf{R}-\mathbf{d_y},\sigma} \psi^{\dagger}_{\mathbf{R},\sigma}) ] +h.c. \\
\end{array}  
\end{equation}
Here, $H_{\Delta_s}$ and $H_{\Delta_p}$ represent the s-wave and p-wave pairing terms respectively. $\sigma$ is the spin index and $\epsilon_{\uparrow, \downarrow}= \pm 1 $.
 In the momentum space and assuming periodic boundary conditions, the pairing terms can be written as:
\begin{equation}
\begin{array}{ll}
H_{\Delta_s} = &  \Delta_s \sum_{\vec{k}} [\psi^{\dagger}_{\vec{k}, \uparrow} \psi^{\dagger}_{-\vec{k}, \downarrow} + h.c.]\\
H_{\Delta_p} = & \Delta_p \sum_{\vec{k}} [( \sin k_y + i \sin k_x ) \psi^{\dagger}_{\vec{k}, \uparrow} \psi^{\dagger}_{-\vec{k}, \uparrow} \\ & - (\sin k_y - i \sin k_x ) \psi^{\dagger}_{\vec{k}, \downarrow} \psi^{\dagger}_{-\vec{k}, \downarrow} + h.c. ]\\
\end{array}  
\end{equation}

\begin{figure}
\includegraphics[width=3.4in]{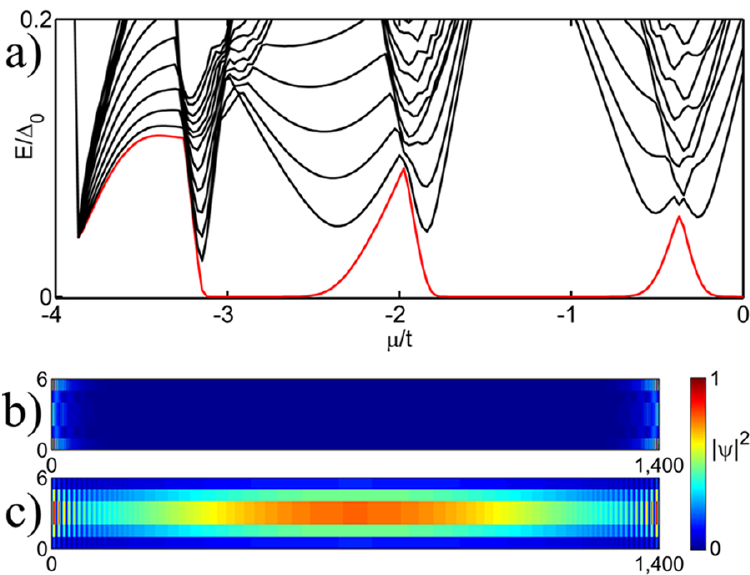}
\caption{\label{SF2} a) Energy spectrum of a wire with $d_{x^2 - y^2}$-wave, s-wave, p-wave pairing and Rashba spin-orbit coupling. $\Delta_s = \Delta_p = 0.2 \Delta_0 $ is assumed.  The parameters used are the same as the parameters in Fig.4a of the main text. b) A plot of the sum of the amplitudes of the ground state wavefunctions of the wire $|\Psi|^2=|\Psi_{1}|^2+|\Psi_{2}|^2$ in the topologically non-trivial regime. Here, $\Psi_{1}$ and $\Psi_{2}$ are the Kramers pair. It is evident that the wavefuntions are localized at the ends c) A plot of the sum of the amplitudes of the ground state wavefunctions of the wire in the topologically trivial regime. The wavefunctions are predominately in the bulk.}
\end{figure}

Since both of these pairing terms preserve time-reversal symmetry and particle-hole symmetry, adding these terms to $H_{q1D}$ of the main text does not change the symmetry class of the Hamiltonian. Therefore, we expect that the presence of Majorana end states is not affected by adding the s-wave and p-wave pairing, as long as these terms do not close the energy gap.  In Fig.\ref{SF2}a, the energy spectrum of a quasi-one dimensional wire with $d_{x^2-y^2}$-wave as well as the s-wave and p-wave pairings are shown. The ground state wave functions in the topologically non-trivial and trivial regimes are shown in Fig.\ref{SF2}b and Fig.\ref{SF2}c respectively. It is evident that the Majorana end states are robust in the presence of the s-wave and p-wave pairing terms. In Fig.\ref{SF2}, $\Delta_s=\Delta_p=0.2 \Delta_0$ is assumed where $\Delta_0$ is the $d_{x^2-y^2}$-wave pairing amplitude.

\section{Multi-layer systems and application to $\text{CeCoIn}_5$ }
In the main text and in the above sections, we show that quasi-one dimensional $d_{x^2-y^2}$-wave superconductors with spin-orbit coupling terms are DIII class TSs in the absence of an external magnetic field. In this section, we show that multi-layers of quasi-one dimensional d-wave superconductors coupled through inter-layer tunneling can be a TS. In particular, we study a system with spatially modulated Rashba terms in which the strength of the Rashba terms in the top layer and bottom layer are non-zero but the Rashba terms in the middle layers are zero. We will argue below that such a model describes a multi-layer $d_{x^2-y^2}$-wave superconductor $\text{CeCoIn}_5$.

The Hamiltonian of a multi-layer  $d_{x^2-y^2}$-wave superconductor with spatially modulated Rashba terms and inter-layer hoppings can be written as: $H_{T}= \sum_{m=1}^{N} H_{m}+ H_{t_{z}}$ where $H_{m}$ is the Hamiltonian for each individual layer and $H_{t_z}$ describes inter-layer hoppings. Here, $m$ is the layer label, $N$ is the total number of layers. Explicitly, $H_{m}$ and $H_{t_z}$ can be written as: 
\begin{equation}
\begin{array}{ll}
H_{m} = & H_{tm} + H_{SOm}+H_{SCm}, \\
H_{tm}= & \sum_{\mathbf{R},\mathbf{d}, \alpha} -t(\psi^{\dagger}_{\mathbf{R+d},\alpha,m}\psi_{\mathbf{R}, \alpha,m}+h.c.) \\ & -\mu \psi^{\dagger}_{\mathbf{R},\alpha,m}\psi_{\mathbf{R}, \alpha,m}\\
H_{SOm}= & \sum_{\mathbf{R,d},\alpha,\beta}-\frac{i{\alpha_{R}}_{m} }{2} \psi^{\dagger}_{\mathbf{R+d},\alpha,m} \hat{\mathbf{z}}\cdot(\vec{\sigma}_{\alpha \beta}\times \mathbf{d})\psi_{\mathbf{R},\beta,m} \\ & +h.c.\\
H_{SCm}= & \sum_{\mathbf{R}}\frac{\Delta_{0}}{2}[ (\psi^{\dagger}_{\mathbf{R+d_{x}},\uparrow,m}\psi^{\dagger}_{\mathbf{R},\downarrow,m}-\psi^{\dagger}_{\mathbf{R+d_{x}},\downarrow,m}\psi^{\dagger}_{\mathbf{R},\uparrow,m}) \\ 
&- (\psi^{\dagger}_{\mathbf{R+d_{y}},\uparrow,m}\psi^{\dagger}_{\mathbf{R},\downarrow,m}-\psi^{\dagger}_{\mathbf{R+d_{y}},\downarrow,m}\psi^{\dagger}_{\mathbf{R},\uparrow,m}) +h.c. ] 
\end{array}  \label{HTm}
\end{equation}
\begin{equation}
\begin{array}{ll}
H_{t_z} =& \sum_{\mathbf{R},\alpha,\langle m, m' \rangle} -t_z (\psi^{\dagger}_{\mathbf{R},\alpha,m}\psi_{\mathbf{R}, \alpha,m'}+h.c.)
\end{array}
\end{equation}

Here, $\psi_{\mathbf{R},\alpha,m}$ represents a fermion annihilation operator at position $\mathbf{R}$ and spin $\alpha$ on layer $m$. $H_{tm}$, $H_{SOm}$, and $H_{SCm}$ are the kinetic, spin-orbit coupling and the superconducting pairing terms respectively. It was first pointed out in Ref.\onlinecite{Sigrist} that such a Hamiltonian, with the possibility of including small s-wave and p-wave pairing terms, describes multi-layers of $\text{CeCoIn}_5$ with spatially modulated Rashba terms.

$\text{CeCoIn}_5$ is a layered $d_{x^2-y^2}$-wave heavy fermion superconductor. Even though many heavy fermion superconductors  break inversion symmetry in the bulk and are non-centrosymmetric superconductors, bulk $\text{CeCoIn}_5$ respects inversion symmetry and it is \emph{not} a non-centrosymmetric superconductor. 

\begin{figure}
\includegraphics[width=3.2in]{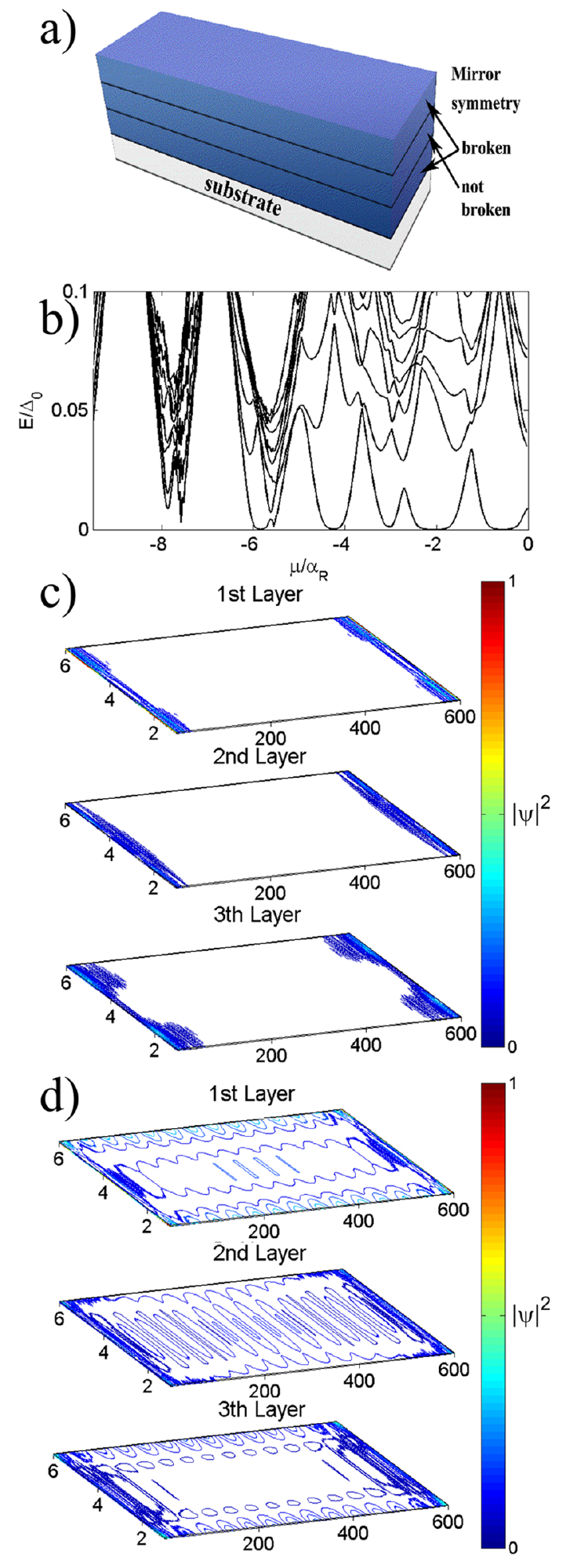}
\caption{\label{SF3}a)A systematic picture of CeCoIn$_5$ modeled as a tri-layers system on a substrate. The top and bottom layers break the mirror symmetry with respect to the z-axis locally  b) The energy spectrum of a finite system with open boundary conditions. The parameters used are $\Delta_0=2, t = 9\Delta_0,\alpha_R = 2\Delta_0$ and $t_z =4.5\Delta_0$  c) The ground state wavefunctions in the topologically trivial regime at $\mu=-2\alpha_R$. d) The ground state wavefunctions in the topologically non-trivial regime at $\mu=-1.25\alpha_R$. }
\end{figure}

However, for multi-layers of $\text{CeCoIn}_5$  sandwitched between the vacuum and a substrate, the top and bottom layers, which are in contact with the vacuum and with the substrate respectively, break the mirror symmetry with respect to the $z$-axis locally. This is illustrated in Fig.\ref{SF3}a. Due to the breaking of mirror symmetry with respect to the $z$-axis on the surface layers and the strong spin-orbit coupling of the $\text{Ce}$ atoms, the surface layers acquire Rashba terms [\onlinecite{Sigrist}]. On the other hand, mirror symmetry of the inner layers is not broken, the inner layers have no Rashba type spin-orbit coupling terms. As a result, this system has spatially modulated Rashba spin-orbit coupling terms [\onlinecite{Sigrist}]. In the case of Fig.\ref{SF3}a, there are only three layers. One may assume that the Rashba terms in the top and bottom layers are non-zero but the Rashba terms of the middle layer vanishes. 

In the following, we consider a three layer system with $\alpha_{R_m} = (\alpha_R, 0, -\alpha_R/2)$.  $\alpha_{R_m}$ is spatially different for different layers because the difference between the vacuum and the substrate breaks the global inversion symmetry. Importantly, we consider a quasi-one dimensional geometry such that the bulk spectrum is gapped as in the single layer case. The energy spectrum of a finite system with open boundary conditions is shown in Fig.\ref{SF3}b. It is evident that zero energy Majorana modes exist. The ground state wavefunctions in the topologically trivial and non-trivial regimes are plotted in Fig.\ref{SF3}c and Fig.\ref{SF3}d respectively. It is evident that the Majorana end states exist in the topologically non-trivial regime. 

In Ref.\onlinecite{Sigrist}, the authors considered a system with global inversion symmetry in which layers of $\text{CeCoIn}_5$ are sandwitched between identical $\text{YbCoIn}_5$ layers [\onlinecite{Matsuda}]. For example,  in the case of a three layer system, $\alpha_{R_m} = (\alpha_R, 0, -\alpha_R)$ is chosen in Ref.\onlinecite{Sigrist}  such that global inversion symmetry is preserved. However, it can be shown  that such a system is topologically trivial. 

\section{The Importance of the quasi-one dimensional geometry}
It is well known that two-dimensional $d_{x^2-y^2}$-wave superconductors are nodal and the pairing gap vanishes along the nodal directions due to the fact that $|k_x|$ can be equal to$ |k_y|$ at the Fermi energy.  However, in a quasi-one dimensional wire, $k_y$ is quantized and it is possible that $|k_y| \neq |k_x|$ for all $k_x$ at the Fermi energy. In this case, the pairing terms do not vanish at the Fermi energy and the system is fully gapped. 

\begin{figure}

\includegraphics[width=3.4in]{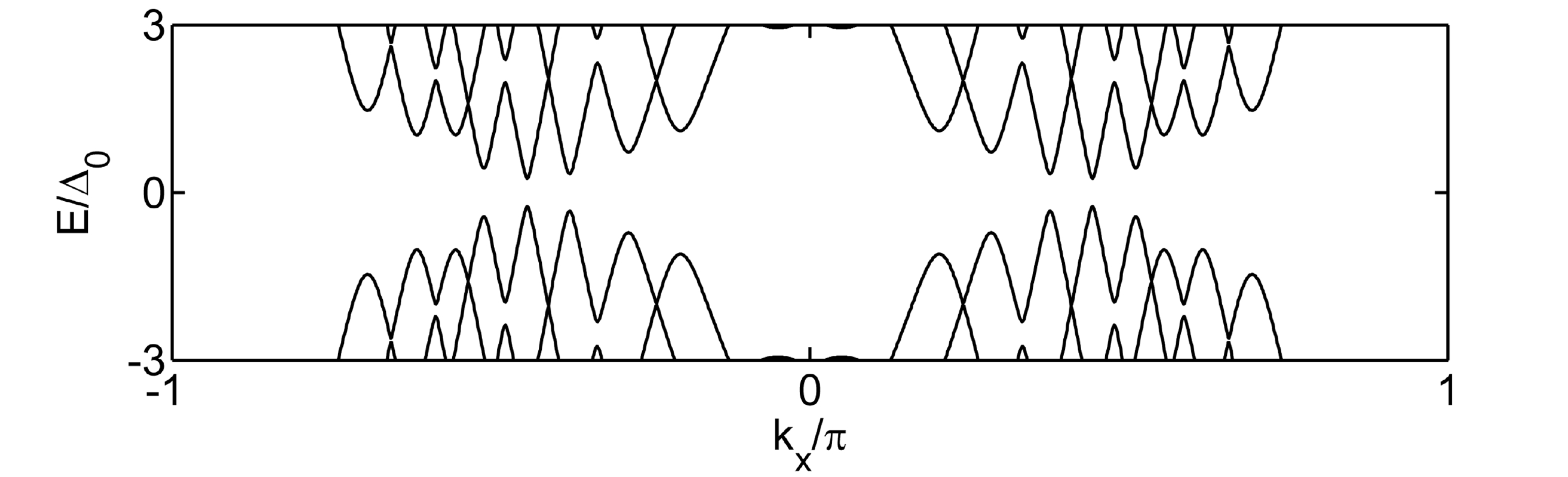}
\caption{\label{SF4}  The bulk band structure for a system with periodic boundary conditions in the x-direction. There are no Majorana end states due to the peridic boundary conditions. The spectrum is fully gapped in the topologically non-trivial regime. The system can undergo a quantum phase transition from topologically non-trivial to topologically trivial phase or vise versa when the bulk gap is closed by tuning the chemical potential. }
\end{figure}

To show that the bulk energy spectrum is gapped, we study a system which has periodic boundary condition in the $x$-direction and open boundary condition in the $y$-direction. The parameters are chosen to be the same as the finite size system in Fig.\ref{Q1D}a of the main text. Due to the annular geometry, there are no end states. The bulk excitation energy of the system versus $k_x$ at $\mu=-t$ is shown in Fig.\ref{SF4}. At $\mu=-t$, the system is in the topologically non-trivial regime. It is evident from Fig.\ref{SF4} that the spectrum is fully gapped.

\section{Majorana fermions in a nodal superconductor}
In the main text, the possibility of realizing Majorana fermions in intrinsic quasi-one dimensional $d_{x^2-y^2}$-wave superconductors are discussed. Another possible way of creating $d_{x^2-y^2}$-wave pairing on a wire with Rashaba coupling is to induce $d_{x^2-y^2}$-wave superconductivity on the wire through proximity effect. Inducing d-wave pairing on wires can be experimentally challenging. In this section, we only discuss how Majorana fermion end states can survive  in the presence of a nodal background. It is interesting to note that in this situation, the Majorana fermions on the wire can couple to the nodal fermions in the d-wave superconductor and it is not obvious that Majorana fermions can survive in the presence of nodal fermions. In this section, we show that Majorana end states on the wire can still survive, even though part of the Majorana wavefunction can leak into the d-wave superconductor.

\begin{figure}
\includegraphics[width=3.6in]{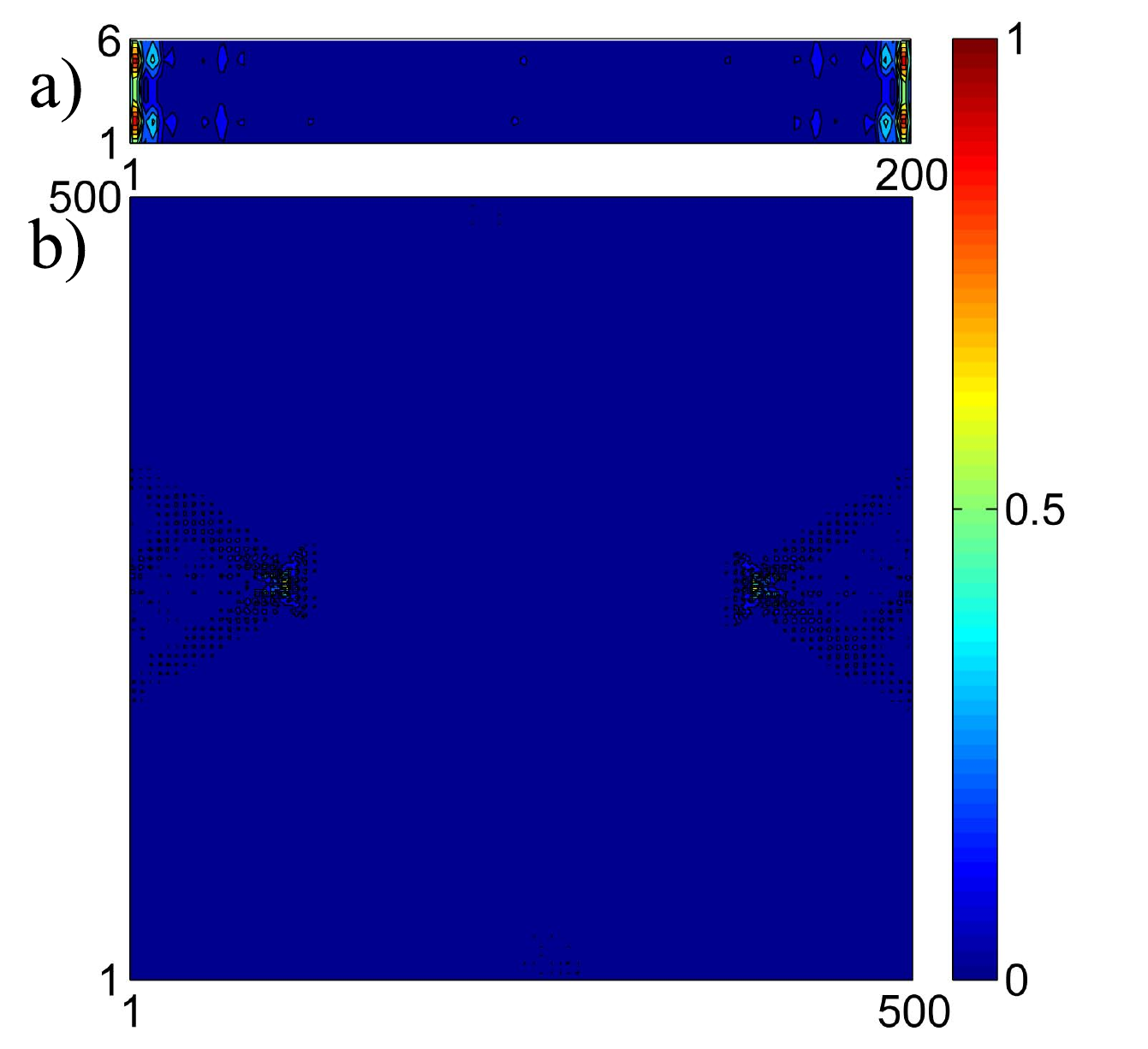}
\caption{\label{SF5} The sum of the amplitude of the two ground state wavefunctions, $|\psi|^2= |\psi_{1}|^2+|\psi_{2}|^2 $, of a system which consists of a quasi-one dimensional superconducting wire placed on top of a two-dimensional $d_{x^2-y^2}$-wave superconductor.  The parameters are given in the text. a) The groud state wavefunction on the quasi-one dimensional wire when the wire is in the topologically non-trivial regime. It is evident that the ground state wavefunction is localized at the ends of the wire. b) The ground state wavefunctions leak into the bulk of the background d-wave superconductor. The wavefunctions leak into the nodal directions in which $|k_x| = |k_y|$.}
\end{figure}

To show this, we couple a quasi-one dimensional wire, which is described by Hamiltonian $H_{q1D}$ in Eq.11 of the main text in the absence of the superconducting pairing terms, to a nodal $d_{x^2-y^2}$-wave superconductor. The length of the wire is 300 (in units of lattice spacing) and the width of is 6. The d-wave superconductor has length 500 and width 500 and the pairing amplitude on the superconductor is $\Delta_0$. The parameters are chosen such that the dimensions of the d-wave superconductor is much larger than the coherence length. It is important to note that the pairing amplitude on the wire is zero. However, due to the coupling between the metal wire and the superconductor, an effective pairing is induced on the wire through proximity effect. The Rashba coupling strength on the wire is chosen to be $\alpha_R = 2 \Delta_0$ and $0$ on the d-wave superconductor. The wire is placed on top of the center of the d-wave superconductor. The hopping amplitudes on the wire and on the d-wave superconductor are chosen to be $t=5\Delta_0$. Each site on the wire is coupled to the site underneath it on the d-wave superconductor through hopping $t_z=t/2$. 

The whole system is then diagonalized numerically and the ground state wavefunction of the whole system is plotted in Fig.\ref{SF5}. Fig.\ref{SF5}a shows the wavefunction on the wire and Fig.\ref{SF5}b shows the wavefunction on the d-wave superconductor. At $\mu=-4t+6$, the wire is topologically non-trivial. It is evident that the ground state wavefunction is predominantly localized at the ends of the wire from Fig.\ref{SF5}a. From Fig.\ref{SF5}b, one can see that part of the Majorana wavefunction leaks into the bulk of the d-wave superconductor. It is important to note that the wavefunction leaks into the nodal directions in which $|k_x| = |k_y|$. 

Due to the presence of the gapless nodal directions, the Majorana fermions are no longer fully localized at the ends of the wire. However, since the wavefunctions can leak into the nodal directions only, the Majorana fermions at the two ends of the wire do not couple to each other directly in the x-direction as the x-direction is fully gapped. As a result, even though the energy of the Majorana modes is increased because of the small overlap of the wavefunction, this energy increase is small and the Majorana nature of the end states is well preserved.

\end{document}